\documentclass[a4paper,11pt]{article}
\usepackage{jheppub} 
\usepackage[T1]{fontenc} 
\usepackage{graphicx}
\usepackage{bm}
\usepackage{subfigure}
\usepackage[mathlines]{lineno}
\usepackage{tabularx}                         %
\usepackage{booktabs}
\usepackage{multirow}
\usepackage{graphicx}                         %
\usepackage{amsmath}                          %
\usepackage{amssymb}                          %
\usepackage{units}        %
\usepackage{siunitx}
\usepackage{gensymb}
\usepackage[normalem]{ulem}
\usepackage{xcolor}       %
\newcommand{\nuebar}{$\overline{\nu}_{e}$}
\newcommand*\sinsq[2]{\ensuremath{\sin^2\theta_{#1#2}}}
\newcommand*\dmsq[2]{\ensuremath{\Delta m^2_{#1#2}}}
\newcommand\ktonGWyear{${\rm kton}\cdot {\rm GW}\cdot {\rm year}$}

\title{Towards a Sub-percent Precision Measurement of \ensuremath{\sin^2\theta_{13}} with Reactor Antineutrinos}
\author[a,b]{Jinnan Zhang}
\author[a,b]{and Jun Cao}
\affiliation[a]{Institute~of~High~Energy~Physics, Chinese Academy of Sciences, Beijing 100049, China}
\affiliation[b]{School of Physical Sciences, University~of~Chinese~Academy~of~Sciences, Beijing 100049, China}
\emailAdd{zhangjinnan@ihep.ac.cn}
\emailAdd{caoj@ihep.ac.cn}

\abstract{
  Measuring the neutrino mixing parameter \ensuremath{\sin^2\theta_{13}} to the sub-percent precision level could be necessary in the next ten years for the precision unitary test of the PMNS matrix. In this work, we discuss the possibility of such a measurement with reactor antineutrinos. We find that a single liquid scintillator detector on a reasonable scale could achieve the goal. We propose to install a detector of $\sim10$\% energy resolution at about 2.0~km from the reactors with a JUNO-like overburden. The integrated luminosity requirement is about 150~${\rm kton}\cdot {\rm GW}\cdot {\rm year}$, corresponding to 4 years' operation of a 4~kton detector near a reactor complex of 9.2 GW thermal power like Taishan reactor. 
  Unlike the previous $\theta_{13}$ experiments with identical near and far detectors, which can suppress the systematics especially the rate uncertainty by the near-far relative measurement and the optimal baseline is at the first oscillation maximum of about 1.8~km, a single-detector measurement prefers to offset the baseline from the oscillation maximum.
  At low statistics $\lesssim 10$~${\rm kton}\cdot {\rm GW}\cdot {\rm year}$, the rate uncertainty dominates the systematics, and the optimal baseline is about 1.3~km. At higher statistics, the spectral shape uncertainty becomes dominant, and the optimal baseline shifts to about 2.0~km. 
  The optimal baseline keeps being $\sim 2.0$~km for an integrated luminosity up to $10^6$~${\rm kton}\cdot {\rm GW}\cdot {\rm year}$.
  Impacts of other factors on the precision \ensuremath{\sin^2\theta_{13}} measurement are also discussed. We have assumed that the TAO experiment will improve our understanding of the spectral shape uncertainty, which gives the highest precision measurement of reactor antineutrino spectrum for neutrino energy in the range of 3--6~MeV. We find that the optimal baseline is $\sim 2.9$~km with a flat input spectral shape uncertainty provided by the future summation or conversion methods' prediction. The shape uncertainty would be the bottleneck of the \ensuremath{\sin^2\theta_{13}} precision measurement. The \ensuremath{\sin^2\theta_{13}} precision is not sensitive to the detector energy resolution and the precision of other oscillation parameters.
}

\begin{document}

\maketitle


\section{Introduction}
\label{sec:introduction}
Since first detected in 1956 by Reines and Cowan at the Savannah River reactor, neutrinos have played an inspiring role in particle physics. Plenty of experiments prove that there are three flavors of neutrinos, and they are massive. Neutrinos are created via electroweak interactions in flavor states ${\nu}_e$, ${\nu}_\mu$, and ${\nu}_\tau$. The neutrino oscillation phenomenon reveals that the three mass eigenstates $\nu_{1,2,3}$ with masses $m_{1,2,3}$ are non-degenerate from the three flavor eigenstates. The Pontecorvo-Maki-Nakagawa-Sakata (PMNS)~\cite{Pontecorvo:1967fh, Maki:1962mu} matrix $U$ is proposed to describe the mixing among massive neutrino states, $\nu_\alpha=U_{\alpha i}\nu_i$, where $\alpha$ indexes flavor states and $i$ indexes mass states. The mixing matrix can be parameterized with three mixing angles, $\theta_{12}$, $\theta_{13}$, and $\theta_{23}$, plus a CP violation phase $\delta_{\rm CP}$. If neutrinos are Majorana fermions, there will be two additional phases irrelevant to the neutrino oscillation. See Ref.~\cite{Athar:2021xsd} for comprehensive reviews and perspectives of neutrino physics.

Table~\ref{tab:current:knowledge} summarizes the current precision estimation of the oscillation parameters and the dominant types of experiments, taken from PDG2020~\cite{ParticleDataGroup:2020ssz}, as well as the projected precision in the near future.
Most of them are determined with precision within a few percent except for $\delta_{\rm CP}$, which is expected to be determined by the next generation neutrino experiments. The measurement of \sinsq12 and \dmsq21 are currently dominated by the solar neutrino experiments SNO~\cite{SNO:2011hxd} and Super-Kamiokande~\cite{nakajima2020recent,Super-Kamiokande:2016yck}, and the long-baseline reactor neutrino experiment KamLAND~\cite{KamLAND:2013rgu}. The accelerator and atmospheric neutrino experiments explore oscillation physics via the same oscillation channels and are sensitive to the same parameters, \dmsq32/\dmsq31, \sinsq23, and $\delta_{\rm CP}$. The dominant measurements are from the IceCube~\cite{IceCube:2017lak}, MINOS~\cite{MINOS:2014rjg}, NOvA~\cite{NOvA:2019cyt}, Super-Kamiokande~\cite{Super-Kamiokande:2017yvm}, and T2K~\cite{T2K:2019bcf} experiments. The precision of the smallest mixing angle \sinsq13 is dominated by the reactor experiments with baselines at $\sim\mathcal{O}(1)$~km, including the Daya Bay~\cite{DayaBay:2018yms}, Double Chooz~\cite{DoubleChooz:2019qbj}, and RENO~\cite{RENO:2019otc} experiments.
\begin{table}[htbp]
  \centering
  \begin{tabular}{lccclc}
    \toprule
    Parameter                                                            & Ordering & Value           & 1$\sigma$ (\%) & Dominant Exps.                       & Prospect (\%, years)                         \\
    \midrule
    $\dfrac{\Delta m^2_{21}}{10^{-5}\ {\rm eV}^2}$                       & NO, IO   & 7.53$\pm$0.18   & 2.4            & Rea., Sol.                           & (0.3, 6) \cite{JUNO:2022pmop}                \\[1.1ex]
    \midrule
    \multirow{2}[0]{*}{$\dfrac{|\Delta m^2_{32}|}{10^{-3}\ {\rm eV}^2}$} & NO       & 2.453$\pm$0.034 & 1.4            & \multirow{2}[0]{*}{Acc., Atm., Rea.} & (0.2, 6) \cite{JUNO:2022pmop}                \\
                                                                         & IO       & $2.546\pm$0.037 & 1.5            &                                                                                     \\
    \midrule
    $\sin^2 \theta_{12}$                                                 & NO, IO   & 0.307$\pm$0.013 & 4.2            & Sol., Rea.                           & (0.5, 6) \cite{JUNO:2022pmop}                \\
    \midrule
    $\dfrac{\sin^2 \theta_{13}}{10^{-2}}$                                & NO, IO   & 2.18$\pm$0.07   & 3.2            & Rea.                                 & (2.9, - )\cite{kam_biu_luk_2022_6683712}     \\
    \midrule
    \multirow{2}[0]{*}{$\sin^2 \theta_{23}$}                             & NO       & 0.545$\pm$0.021 & 3.9            & \multirow{2}[0]{*}{Acc., Atm.}       & (0.7--3.4, 10)                               \\
                                                                         & IO       & 0.547$\pm$0.021 & 3.8            &                                      & \cite{DUNE:2020lwj,Hyper-Kamiokande:2018ofw} \\
    \midrule
    $\delta_{\rm CP}$/$\pi$                                              & NO, IO   & 1.36$\pm$0.17   & 12.5           & Acc., Atm.                           & TBD                                          \\
    \bottomrule
  \end{tabular}%
  \caption{Current relative precision estimation of the oscillation parameters from PDG2020~\cite{ParticleDataGroup:2020ssz} and the projected precision in the near future with corresponding data taking time. The 1$\sigma$ relative uncertainties and the dominant types of experiments (Exps.) are also listed. The reactor experiments (Rea.) measure \sinsq13 and \dmsq21 with the highest precision. The solar experiments (Sol.) contribute a lot to the \sinsq12 and \dmsq21 determination. The accelerator (Acc.) and atmospheric experiments (Atm.) dominate the measurements of \sinsq23, \dmsq32, and $\delta_{\rm CP}$. The estimation for both normal ordering (NO) and inverted ordering (IO) are presented.
  }
  \label{tab:current:knowledge}%
\end{table}%

Towards the sub-percent precision measurements of the oscillation parameters, Table~\ref{tab:current:knowledge} also lists the prospects in the next ten years. The increasing data volume at the NOvA and T2K experiments will further improve the precision of \dmsq31/\dmsq32 to about 1\%~\cite{NOvA:2004blv, T2K:2011qtm}. The under-construction medium baseline reactor neutrino experiment JUNO will start operation in 2023 and can measure \sinsq12, \dmsq21, and \dmsq31/\dmsq32 to the sub-percent precision level within one year~\cite{JUNO:2022hxd, JUNO:2022pmop}. The next-generation experiments DUNE~\cite{DUNE:2020lwj} and Hyper-Kamiokande~\cite{Hyper-Kamiokande:2018ofw} are designed to determine the CP-violation phase $\delta_{\rm CP}$, the octant of \sinsq23, and the mass ordering. After ten years of data taking, they can measure \sinsq23 to the precision from sub-percent to $\sim 3.4$\%, depending on the true octant. DUNE and Hyper-Kamiokande will also measure \dmsq32/\dmsq31 to the sub-percent precision level with channels other than JUNO. The DUNE measurement of \sinsq13 can also approach the precision of Daya Bay with high exposure~\cite{DUNE:2020lwj}.
In the next ten years, we can expect sub-percent precision measurements for most oscillation parameters, \sinsq12, \sinsq23, \dmsq21, and \dmsq32/\dmsq31.
However, for \sinsq13, there are no more experiments under construction for more precise measurement. The only under active exploration project is the SuperChooz~\cite{anatael_cabrera_2022_7504162}, which was preliminary started in 2018 and officialized in 2022.
The Daya Bay experiment is shut down in 2020. The precision of \sinsq13 with the full dataset of neutron-capture on Gadolinium (nGd) is expected to be about 2.7\%~\cite{DayaBay:2018yms}. In the recent Neutrino 2022 conference, the Daya Bay experiment's latest precision of \sinsq13 using full nGd data is about 2.9\%~\cite{kam_biu_luk_2022_6683712}.

Measuring \sinsq13 to a sub-percent precision level would be important for various research fields, including particle physics, astrophysics, and cosmology~\cite{Athar:2021xsd,JUNO:2022pmop}.
For example, it will enable more stringent tests of the standard 3 flavor neutrino mixing picture, such as probing the unitarity of the PMNS matrix~\cite{Antusch:2006vwa,Parke:2015goa,Fong:2016yyh,Blennow:2016jkn,Li:2018jgd,Ellis:2020hus} and exploring the physics beyond the standard model.
The sub-percent precision knowledge of the leptonic mixing matrix may help reveal its fundamental structure and provide important clues for identifying the theoretical mechanisms behind neutrino mass and mixing generation~\cite{King:2019gif}.
It can also reduce the parameter space for searching the leptonic CP violation~\cite{DUNE:2015lol,Hyper-KamiokandeProto-:2015xww} and neutrinoless double beta decay~\cite{Dueck:2011hu,Ge:2015bfa,Cao:2019hli}.
Finally, with high-precision oscillation parameters, we can employ the neutrinos as a more reliable messenger in probing the deep interiors of astrophysical objects such as the Sun, supernovae, and Earth.

In this work, we discuss the possibility of measuring \sinsq13 to the sub-percent precision level with reactor antineutrinos via the Inverse Beta Decay (IBD) interaction. The electron antineutrinos (\nuebar)~from nuclear reactors are a very powerful source for measuring the $\theta_{13}$ mixing angle. We could get large statistics and predict the energy spectrum precisely.
The analysis methods used in this work are similar to the classical reactor neutrino analyses~\cite{Huber:2003pm,DayaBay:2007fgu,JUNO:2015zny}. There are some discussions about \sinsq13 precise measurement, both at the time before determined its value~\cite{Huber:2003pm,Huber:2006vr} and after that~\cite{Cabrera:2019xkf}.
Unlike their proposal to build multiple detectors for suppressing the reactor-related uncertainties, we find a single detector can also measure \sinsq13 to the sub-percent precision.
The same single detector methodology is implemented in the Double Chooz experiment before its near detector is online~\cite{DoubleChooz:2011ymz,DoubleChooz:2012gmf,DoubleChooz:2014kuw}, although the \sinsq13 relative precision is poor due to its setup.
This paper is organized as follows. We present the survival probability calculation for \nuebar, the observed spectrum prediction, and the analysis strategy in Sec.~\ref{sec:detection-statistics}. In Sec.~\ref{sec:sensitivity}, the numerical results show that, by installing a single 4~kton detector of $\sim 10$\% energy resolution at the baseline of $\sim 2.0$~km from a reactor complex like Taishan reactor, the experiment could measure \sinsq13 to the sub-percent precision level within about 4~years. We also study the impact of different factors that may increase or decrease the sensitivity in that section. Finally, the summary and conclusions are posted in Sec.~\ref{sec:conclusion}.

\section{Reactor antineutrino detection and statistical analysis}
\label{sec:detection-statistics}
The IBD interaction, $\bar{\nu}_e+p\to n+e^+$, is the typical channel to detect the \nuebar~in the few-MeV range with liquid scintillator (LS) detectors for its large cross-section. In this reaction, the electron antineutrino interacts with a proton ($p$) in the LS, creating a positron ($e^+$) and a neutron ($n$). The $e^+$ takes most energy of the original neutrino and quickly deposits its energy and annihilates into gammas, giving a prompt signal. Thus, the experiment can extract neutrino oscillation parameters by looking at the observed positron energy spectrum. While the neutron is thermalized in the detector and captured by a nucleus, it produces a delayed signal. The time, space, and energy correlation between the prompt and delayed signals are powerful to suppress the backgrounds. This section presents the approach to predicting the visible energy spectrum of the reactor antineutrinos with IBD reaction. Then we introduce the statistical method to calculate the \sinsq13 sensitivity. We also carefully estimate the systematic uncertainties based on the experiences of previous and current experiments.

\subsection{Reactor antineutrino spectrum prediction}
\label{sec:detection-statistics:detection}
The commercial nuclear power plants are ``free'' and powerful artificial \nuebar~source for measuring \sinsq13, which generate electron antineutrinos via subsequent $\beta$ decays of the fission products of mainly four isotopes, $^{235}$U, $^{238}$U, $^{239}$Pu, and $^{241}$Pu. After creation, the \nuebar~propagates in mass eigenstates on the way to the detector and then is detected with the weak eigenstate \nuebar. Under the assumption of three flavor mixing, the survival probability ${\cal P}(\bar{\nu}_e\to\bar{\nu}_e)$ can be expressed as,

\begin{equation}
  \begin{aligned}
    {\cal P}  (  \bar{\nu}_e\to\bar{\nu}_e)  = & 1  -\sin^22\theta_{13}(\cos^2\theta_{12}\sin^2\Delta_{31}+\sin^2\theta_{12}\sin^2\Delta_{32}) \\
                                               & -\cos^4\theta_{13}\sin^22\theta_{12}\sin^2\Delta_{21},
  \end{aligned}
  \label{eq:probability:e2e}
\end{equation}
where $\Delta_{ji}=\Delta m_{ji}^2L/(4E_{\bar{\nu}})$, L is the baseline and $E_{\bar{\nu}_e}$ is electron antineutrino energy.
The terrestrial matter effects can influence the oscillation pattern~\cite{Wolfenstein:1977ue, Mikheyev:1985zog}. For a several-kilometer baseline experiment, the matter effects are relatively small. Nonetheless, we include the matter effects in this work~\cite{Lisi:1997yc, Akhmedov:2004ny} with a typical constant matter density $\rho=2.6$~g/$\rm{cm^3}$. The matter effects may distort the survival probability up to relatively 0.2\% with negligible uncertainty.

The visible prompt energy ($E_{\rm prompt}$) spectrum from reactor $r$ at detector $d$ at time $t$ can be predicted as,
\begin{equation}
  \begin{split}
    S_{d,r}&(E_{\rm prompt},t)= \\
    &\left(\epsilon_d\cdot N_{p,d}
    \int{dE_{\bar{\nu}_e} \frac{{\cal P}_{ee}(E_{\bar{\nu}_e},L_{dr})}{ 4\pi L^2_{dr}}\frac{W_r(t)}{\sum_{i} f_{ir}(t) e_i}\sum_i f_{ir}(t)\phi_i(E_{\bar{\nu}_e})}\sigma _{\rm tot}(E_{\bar{\nu}_e})\right)\\
    &\otimes G(E_{e^+},\sigma_d),
  \end{split}
  \label{eq:spec-prediction}
\end{equation}
where $\epsilon_d$ and $N_{p,d}$ are the detection efficiency and the number of target protons of detector $d$, respectively. A 12\% hydrogen fraction from the Daya Bay experiment~\cite{DayaBay:2012aa} is used to calculate the number of target protons with corresponding target mass. ${\cal P}_{ee}(E_{\bar{\nu}_e},L_{dr})$ is the \nuebar~survival probability with energy $E_{\bar{\nu}_e}$ and baseline $L_{dr}$ is the distance from detector $d$ to reactor $r$. $W_r(t)$ and $f_{ir}(t)$ are the thermal power and fission fraction of the reactor $r$ at time t. The nuclear reactors release energy by fission reactions, and $e_i$ and $\phi_i(E_{\bar{\nu}_e})$ are the energy yield and neutrino spectrum per fission of isotope $i$, respectively. In this work,  we set the average fission fraction to be 0.564, 0.076, 0.304, and 0.056~\cite{DayaBay:2021dqj} with the mean energy per fission of 202.36~MeV, 205.99~MeV, 211.12~MeV, and 214.26~MeV~\cite{Kopeikin:2004cn, Ma:2012bm} for $^{235}$U, $^{238}$U, $^{239}$Pu, and $^{241}$Pu, respectively. The Huber-Mueller model ($^{235}$U, $^{239}$Pu, and $^{241}$Pu from Ref.~\cite{Huber:2011wv}, $^{238}$U from Ref.~\cite{Mueller:2011nm}) is a widely used model for calculating the \nuebar~energy spectrum of the isotopes.
The measurements from the reactor neutrino experiments such as Bugey-4~\cite{Declais:1994ma}, Daya Bay~\cite{DayaBay:2018heb}, DANSS~\cite{DANSS:2018fnn}, Double Chooz~\cite{DoubleChooz:2019qbj}, NEOS~\cite{NEOS:2016wee}, Neutrino-4~\cite{NEUTRINO-4:2018huq}, PROSPECT~\cite{PROSPECT:2020sxr}, RENO~\cite{RENO:2020dxd}, and STEREO~\cite{STEREO:2020hup} reveal that both the measured neutrino flux and shape are inconsistent with the Huber-Mueller model. Nevertheless, We find that the measured antineutrino flux and spectrum from the Daya Bay experiment~\cite{DayaBay:2021dqj} and the Huber-Mueller flux model give consistent \sinsq13 sensitivities, which is discussed in detail in Sec.~\ref{sec:sensitivity:RAA}. For simplicity and without losing accuracy, we employ the Huber-Mueller flux model here.

The total cross-section of the IBD reaction $\sigma _{\rm tot}(E_{\bar{\nu}})$ can be precisely calculated~\cite{Vogel:1999zy, Strumia:2003zx, Ricciardi:2022pru}.
The term ``$\otimes G(E_{e^+},\sigma_d)$'' represents the Gaussian smearing processes to take into account the energy resolution of detector $d$ with resolution $\sigma_d$. The energy resolution is not a key factor for \sinsq13 measurement at several kilometers. Thus, we set the energy resolution of the detector to be $10\%/\sqrt{E({\rm MeV})}$. A detector with such resolution would be sufficiently sensitive to \sinsq13 and not too expensive. Nonetheless, we study the impact of the energy resolution in Sec.~\ref{sec:sensitivity:detector}.
Given the prompt energy interval $[E_{\rm prompt,k},E_{\rm prompt,k+1}]$, we can calculate the expected number of signals $T_{d,k}$ as,
\begin{equation}
  T_{d,k}=\int_{E_{\rm prompt,k}}^{E_{\rm prompt,k+1}}dE_{\rm prompt}\int_{\rm t_{DAQ}}dt \sum_r S_{d,r}(E_{\rm prompt},t),
  \label{eq:T_dk}
\end{equation}
where ${\rm t_{DAQ}}$ is the total data taking time.

\subsection{Statistical analysis and systematics}
\label{sec:detection-statistics:statistics}
To extract the \sinsq13 sensitivity of the experiment, we firstly generate a binned Asimov dataset with the nominal setup and approach described above. Then we fit the pseudo data with hypotheses using the Poisson-likelihood $\chi^2$ with nuisance parameters and pull terms to account for the systematic uncertainties~\cite{Stump:2001gu},

\begin{equation}
  \begin{split}
    \chi^2 \equiv 2\sum_{d,k}\left({T_{d,k}({\boldsymbol\theta}, {\boldsymbol\epsilon})-D_{d,k}}+D_{d,k}\ln\frac{D_{d,k}}{T_{d,k}({\boldsymbol\theta}, {\boldsymbol\epsilon})}\right)\\
    +\sum_{s,d,k}\left(\frac{\epsilon_{s,d,k}}{\sigma_{s,d,k}}\right)^2
    +\left({\boldsymbol\epsilon}_{\rm corr}-\overline{\boldsymbol\epsilon}_{\rm corr}\right)^TV_{\epsilon_{\rm corr}}^{-1}\left({\boldsymbol\epsilon}_{\rm corr}-\overline{\boldsymbol\epsilon}_{\rm corr}\right),
  \end{split}
  \label{eq:chisq}
\end{equation}
where $D_{d,k}$ is the event rate in the $k$-th energy bin of detector $d$. $T_{d,k}$ is the predicted value with the oscillation parameters $\boldsymbol{\theta}$ and the nuisance parameters ${\boldsymbol\epsilon}$. $\sigma_{s,d,k}$ in the pull term represents the estimation of the $s$-th systematic uncertainties for the $k$-th energy bin of detector $d$, and $\epsilon_{s,d,k}$ is the corresponding uncorrelated nuisance parameter.
${\boldsymbol\epsilon}_{\rm corr}$, $\overline{\boldsymbol\epsilon}_{\rm corr}$, and $V_{\epsilon_{\rm corr}}$ are the correlated nuisance parameters, their central values, and their covariance matrix, respectively.
In this work, we perform the binned analysis using 320 equal bins for prompt energy from 0.8~MeV to 12~MeV.

To extract the sensitivity of \sinsq13, we minimize the $\chi^2$ defined in Eq.~\eqref{eq:chisq} with respect to all the oscillation parameters and nuisance parameters. For the oscillation parameters, we use the prior central values and uncertainties in Table~\ref{tab:current:knowledge} from PDG2020~\cite{ParticleDataGroup:2020ssz} to constrain \sinsq12, \dmsq21, and \dmsq31.
The \sinsq13 sensitivity weakly depends on these parameters, as shown in Sec.~\ref{sec:sensitivity:oscillation}.
To find the best location and luminosity requirements of the experiment, we defined the 1$\sigma$ precision sensitivity of \sinsq13 as
\begin{equation}
  P_{1\sigma}=\frac{(x^{\rm up}-x^{\rm low})}{2\cdot x_{\rm bft}},
  \label{eq:1sigma:precision}
\end{equation}
where $x^{\rm up}$($x^{\rm low}$) is the upper (lower) bound for parameter $x$ (i.e., \sinsq13) at $1\sigma$ level, i.e., at which the marginalized $\Delta\chi^2(x)\left(\equiv \chi^2(x)-\chi^2_{\rm min}\right)$ is equal to 1. $x_{\rm bft}$ is the best fit value of the parameter.

The systematic uncertainties are especially important for the precision measurement of \sinsq13 with large statistics. Based on the experiences and prospects of reactor experiments Daya Bay~\cite{DayaBay:2016ggj,DayaBay:2018yms,DayaBay:2019fje}, Double Chooz~\cite{DoubleChooz:2020pnv}, KamLAND~\cite{KamLAND:2010fvi}, RENO~\cite{RENO:2019otc}, and JUNO~\cite{JUNO:2022hxd}, here we list the estimation of the systematic uncertainties used in this work in Table~\ref{tab:syst:summary}.
\begin{table}[htbp]
  \centering
  \begin{tabular}{llll}
    \toprule
    Rate                            & Uncertainty (\%)                                              & Shape                                  & Uncertainty (\%)                               \\
    \midrule
    Flux prediction                 & 2.0~\cite{DayaBay:2018heb, DoubleChooz:2019qbj, RENO:2020dxd} & Flux prediction                        & $<$1.0 (2--5~MeV)~\cite{JUNO:2020ijm}          \\
    Fission fraction                & 0.6~\cite{DayaBay:2016ggj,DoubleChooz:2019qbj,RENO:2020dxd}   & Background subtraction                  & 0.1                                            \\
    Thermal power                   & 0.5~\cite{DayaBay:2016ggj,DoubleChooz:2019qbj,RENO:2020dxd}   & Residual nonlinearity                  & 0.3 (event-level)~\cite{JUNO:2020xtj}          \\
    Proton number                   & 0.9~\cite{DayaBay:2016ggj}                                    & Total                                  & With TAO, $\sim 1.0$                           \\
    \cmidrule{3-4}    IBD selection & 2.0                                                           & \multirow{2}[2]{*}{Energy calibration} & \multirow{2}[2]{*}{0.5~\cite{DayaBay:2019fje}} \\
    Total                           & $\sim 3$                                                      &                                        &                                                \\
    \bottomrule
  \end{tabular}%
  \caption{Summary of major systematic uncertainties and their values used in this work. The spectral shape uncertainties are estimated for the bin width of 35~keV.}
  \label{tab:syst:summary}%
\end{table}%
We pack the systematics into several groups: rate uncertainty, spectral shape uncertainty, and energy calibration uncertainty. We estimate their values and assess their impact on \sinsq13 precision measurement sensitivity as follows.
\subsubsection{Overall rate uncertainty}
\label{sec:detection-statistics:statistics:rate}
The overall event rate systematic uncertainty contains all the effects that may affect the normalization of the total event number. We set the value to be 3\% as the quadratic sum of all independent sources. In Eq.~\eqref{eq:chisq}, we assign a nuisance parameter $\epsilon_{{\rm rate}}$ with constraint $\sigma_{\rm rate}$ to take into account this uncertainty. The major source of the rate uncertainty is the predicted number of antineutrinos yielded by the nuclear reactor. The model prediction uncertainty can be constrained by the absolute reactor neutrino flux measurement by the near detectors of the Daya Bay~\cite{DayaBay:2018heb}, Double Chooz~\cite{DoubleChooz:2019qbj}, and RENO~\cite{RENO:2020dxd} experiments, which include the uncertainty of the IBD cross-section, and the overall uncertainty is less than 2\%.
Other short baseline reactor neutrino experiments also provide precise rate measurement of the reactor neutrino flux~\cite{Berryman:2020agd}, such as the Bugey-4~\cite{Declais:1994ma} and Rovno~91~\cite{Kuvshinnikov:1990ry}. In this work, we put a 2\% uncertainty on flux rate prediction.
We also include the rate uncertainties per reactor from the fission fraction (0.6\% in total) and thermal power (0.5\%)~\cite{DayaBay:2016ggj,DoubleChooz:2019qbj,RENO:2020dxd}.

On the detection side, we assume that the uncertainty of the number of target protons would be 0.9\%, close to the Daya Bay experiment~\cite{DayaBay:2016ggj}.
In the pioneer kton LS reactor neutrino experiment KamLAND, the IBD selection efficiency uncertainty is about 2\% and dominated by the fiducial volume cut uncertainty~\cite{KamLAND:2010fvi}. For a future experiment, we anticipate that we will pay more attention to the detector calibration and event reconstruction. These efforts would enable the experiment to control the IBD selection uncertainty to be close to or better than the Daya Bay~\cite{DayaBay:2016ggj} and JUNO~\cite{JUNO:2022pmop} experiment. Nonetheless, we put a 2\% rate uncertainty for the IBD selection.

\subsubsection{Spectral shape uncertainty}
\label{sec:detection-statistics:statistics:shape}
The spectral shape uncertainty refers to the factors that may distort the spectral shape and are uncorrelated from bin to bin. It is essential for high-precision measurement.
The reactor antineutrino flux shape prediction based on the direct measurement of the reactor neutrino experiments~\cite{DayaBay:2021dqj, DoubleChooz:2019qbj, RENO:2020dxd} or from the summation and conversion methods~\cite{Estienne:2019ujo, Hayes:2015yka, Huber:2011wv, Mueller:2011nm} has an uncertainty of 2\% to more than 5\%.
In this work, we set the reactor antineutrino spectral shape uncertainty constrained by the future short-baseline experiment, which is $<$1\% for a bin width of about 35~keV in most signal energy range (2--5~MeV).
In Eq.~\eqref{eq:chisq}, we assign each energy bin a nuisance parameter $\epsilon_{{\rm shape},k}$ with constraint $\sigma_{\rm shape,k}$ to take into account this uncertainty.

The Taishan Antineutrino Observatory (TAO)~\cite{JUNO:2020ijm}, with a ton-level Gadolinium-doped Liquid Scintillator (GdLS) detector at a baseline of 30~m from a reactor core of the Taishan Nuclear Power Plant (NPP), will start operation in 2023 as a satellite experiment of the JUNO experiment. Thanks to the almost full optical coverage with high photon detection efficiency ($>$50\%) Silicon Photomultipliers (SiPMs), the TAO energy resolution is better than 2\% at 1~MeV. After six years of data taking, TAO could measure the reactor spectral shape to the precision of better than 1\% in the prompt energy range of 2--5~MeV, with the bin width of about 35~keV to investigate the possible fine structure in the spectrum.
The reactor \nuebar~flux shape constrained by the direct measurement of the TAO experiment is more precise than the current summation (or ab-initio)~\cite{Hayes:2015yka, Estienne:2019ujo} or the conversion~\cite{Huber:2011wv, Mueller:2011nm} methods; and the shape uncertainty distribution is dominated by the statistics. Differently, the relative shape uncertainty distribution of the summation or conversion model prediction is approximately flat in most of the energy range. We find that the relative shape uncertainty distribution has a large impact on the high precision \sinsq13 measurement, and more details are described in Sec.~\ref{sec:sensitivity:shape_err}.

We also include the uncertainties of other sources that may distort the spectrum's shape.
The antineutrinos from the spent nuclear fuel ($\sim0.3\%$) and non-equilibrium ($\sim0.6\%$) are taken into account using the calculation from Ref.~\cite{DayaBay:2016ssb}, with negligible shape uncertainty. We assign both contributions a 30\% relative rate uncertainty based on the evaluation in Ref.~\cite{Ma:2015lsv} and Ref.~\cite{Mueller:2011nm}.
The background subtraction may also induce spectral shape uncertainty. We do not include the specific background in this work; instead, we assess their impacts within the total spectral shape uncertainty. For a detector with a baseline of several kilometers, we assume the overburden for suppressing the cosmogonic backgrounds is close to the JUNO experiment ($\sim$650~m). We also assume that with a technology similar to JUNO~\cite{JUNO:2021kxb}, the radioactive background would be sufficiently small. We calculate the reactor antineutrino event rate and roughly estimate the rate of the possible backgrounds after the IBD selection with similar criteria as the KamLAND experiment~\cite{KamLAND:2013rgu}. We find that the relative spectrum uncertainty from the background subtraction should be at the level of 0.1\%. The backgrounds considered in the estimations are radioactive background~\cite{JUNO:2021kxb}, cosmogonic background~\cite{DayaBay:2016ggj,KamLAND:2009zwo}, Geo-neutrino and atmospheric neutrinos~\cite{Borexino:2019gps}.

The energy detection uncertainties may also distort the spectrum's shape. Their major sources are energy nonlinearity and relative energy scale.
The spectral shape uncertainty provided by the TAO experiment has taken into account the uncertainty of the LS physics nonlinearity~\cite{JUNO:2020ijm}. With similar LS technology, we assume the LS nonlinearity and its uncertainty of our future experiment are mostly correlated to the TAO experiment.
The residual nonlinearity from the SiPM readout system of the TAO experiment can be neglected in the energy range of 1~MeV to 10~MeV~\cite{Xu:2022mdi}.
With the novel dual calorimetry and dedicated calibration strategy, the JUNO experiment can control the instrumental nonlinearity's uncertainty to a 0.3\% level~\cite{JUNO:2020xtj}. This work assumes the same event-level instrumental nonlinearity uncertainty of 0.3\% as the JUNO experiment and is fully uncorrelated to the TAO experiment. We assign each energy bin a nuisance parameter $\epsilon_{\rm RNL, k}$ to take into account the residual nonlinearity uncertainty. The covariance matrix $V_{\epsilon_{\rm RNL}}$ for these parameters are evaluated by 10,000 toy MC simulations for each experimental setup.

\subsubsection{Energy scale uncertainty}
\label{sec:detection-statistics:statistics:calib}
The relative energy scale uncertainty can be controlled by proper calibration strategy, as verified by the Daya Bay experiment~\cite{DayaBay:2019fje}. This work assumes the relative energy scale calibration uncertainty to be 0.5\%, which is close to the expected value of the coming large detector JUNO~\cite{JUNO:2020xtj}.
To account for this uncertainty, we assign a nuisance parameter $\epsilon_{\rm calib}$ with constraint $\sigma_{\rm calib}$. In predicting the observed energy spectrum of Eq.~\eqref{eq:spec-prediction}, we analytically replace the expected energy $E_{\rm 0}$ by $(1+\epsilon_{\rm calib})\cdot E_{\rm 0}$, on which the energy resolution is defined.

\section{Sub-percent precision measurement of \sinsq13}
\label{sec:sensitivity}
In this work, we quote the \sinsq13 precision measurement sensitivity for different statistics using the integrated luminosity [\ktonGWyear] under the assumption of a single reactor. Take the Daya Bay experiment~\cite{DayaBay:2012fng} as an example. The reactor's thermal power is 17.4~GW. The total target mass of the four far detectors is 80~tons. Thus, the total integrated luminosity is about 12.5~\ktonGWyear~from the start of data taking in 2011 to the shutdown in 2020.

The reactor antineutrino experiments can measure \sinsq13 by installing identical near and far detectors, like the Daya Bay~\cite{DayaBay:2012fng},  Double Chooz~\cite{DoubleChooz:2019qbj}, and RENO~\cite{RENO:2019otc} experiments.
The identical detectors could tremendously reduce the rate systematic uncertainties with the near-far relative measurement.
However, as the statistics increase, the spectral shape systematic uncertainty becomes more important, which can not be suppressed as efficiently as rate uncertainties by the near-far detector strategy.
With the setup presented in Sec.~\ref{sec:detection-statistics}, we numerically calculate the \sinsq13 measurement precision sensitivity for an experiment with identical near and far detectors. We assume only one reactor core and set the detector uncorrelated uncertainty values based on the Daya Bay experiment's experiences~\cite{DayaBay:2016ggj}.
Between the two detectors, we set 0.2\% uncorrelated rate uncertainty, 0.1\% uncorrelated spectral shape uncertainty, and 0.1\% uncorrelated energy calibration uncertainty.
We find the optimal baseline is $\sim$1.8~km, which is well-known and consistent with Ref.~\cite{Huber:2003pm}. With the optimal baseline, we find the identical near and far detectors that could measure \sinsq13 to sub-percent precision level with the luminosity of $\sim$50~\ktonGWyear. Considering the baseline spread in a real reactor complex, the required luminosity for \sinsq13's sub-percent precision measurement could be larger.
At the same time, based on the experiences of the Daya Bay experiment~\cite{DayaBay:2007fgu}, the cost for identical near and far detectors can be used to build a roughly four times larger single far detector.
Besides, if we want to build an experiment with identical near and far detectors, building a kton-level near detector at a very short baseline with no vision of the oscillation effect seems less feasible.
Another strategy is to build non-identical near and far detectors; one example is the recently proposed SuperChooz experiment~\cite{anatael_cabrera_2022_7504162}. For non-identical detectors, the systematical correlation and physics potential is similar to the proposals of using the very short baseline experiments as effective near detectors.

For measuring \sinsq13 to the sub-percent precision efficiently, out of the reasons listed above, 
we discuss only the feasibility of building a single detector alone or employing the TAO experiment as an effective near detector. The latter strategy is usually attractive since we can make use of the under-construction TAO experiment and save the budget from building a new near detector.
With the spectrum prediction and systematic uncertainty estimation in Sec.~\ref{sec:detection-statistics}, we numerically calculate the precision sensitivity of \sinsq13 to find the best choice of the baseline and other configurations. Then, we propose the nominal setup and discuss the impact of various factors on sensitivity.

\subsection{Experiment setup and precision measurement sensitivity}
\label{sec:sensitivity:setup}
\begin{figure}[ht]
  \centering
  \includegraphics[width=\textwidth]{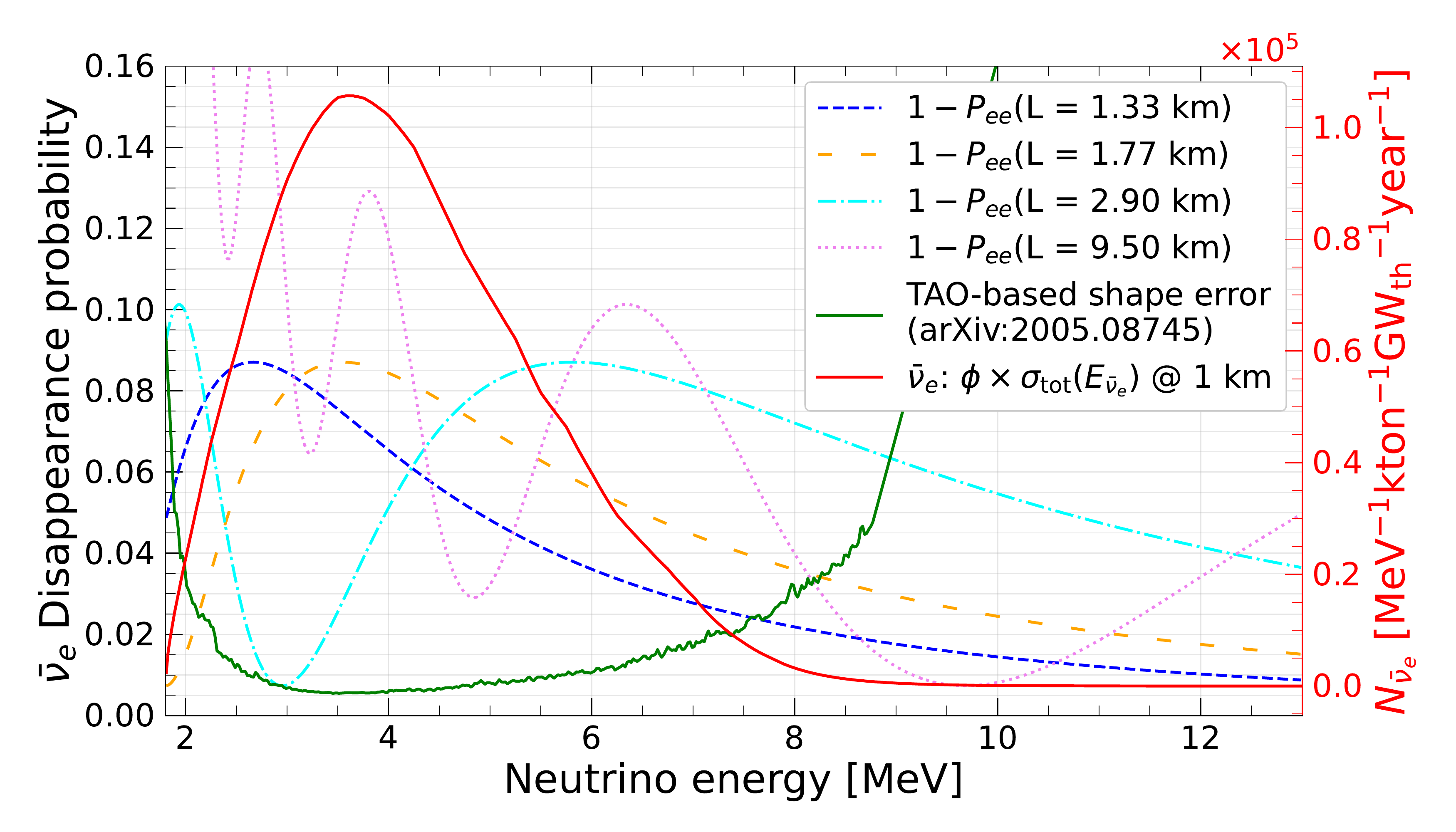}
  \caption{The unoscillated measurable reactor antineutrino energy spectrum $\phi\times \sigma_{\rm tot}(E_{\bar{\nu}_e})$ (flux multiplied by the total IBD cross-section) at 1~km and the \nuebar~disappearance probability $1-P_{ee}$ at different baselines. For a single detector, the rate uncertainty would mimic the oscillation pattern of \sinsq13/\dmsq31 when the oscillation maximum is at the peak of the antineutrino spectrum.
    The relative spectral shape uncertainty constrained by the TAO experiment (green line, same y-scale as left y-axis)~\cite{JUNO:2020ijm} shows that the most precise neutrino energy region is 3--6~MeV.
  }
  \label{fig:flux_oscillation}
\end{figure}

Fig.~\ref{fig:flux_oscillation} shows the unoscillated measurable antineutrino energy spectrum (flux multiplied by the total IBD cross-section) at 1~km, the relative spectral shape uncertainty constrained by the future TAO experiment~\cite{JUNO:2020ijm}, together with the \nuebar~disappearance probability at different baselines.
The \nuebar$\to$\nuebar~disappearance amplitude represents the oscillation parameter \sinsq13; thus, \sinsq13's high precision measurement requires low uncertainties around the disappearance probability peak energy.
The green line in the figure shows that the direct measurement can constrain the shape uncertainty to a sub-percent level for the neutrino energy of 3--6~MeV~\cite{JUNO:2020ijm}. Due to the limit of statistics, however, the relative uncertainties are large at the low and high energy end.
Thus, as discussed in Sec.~\ref{sec:detection-statistics:statistics}, if the future summation or conversion methods could predict the reactor antineutrino flux to be consistent with the measurement of the TAO experiment, the spectral shape uncertainty of their combined prediction would be better than 1\%. Therefore, hereafter we set a 1\% spectral shape relative uncertainty to study the optimal baseline of a future \sinsq13 measurement experiment.

\begin{figure}[htbp]
  \centering
  \includegraphics[width=\textwidth]{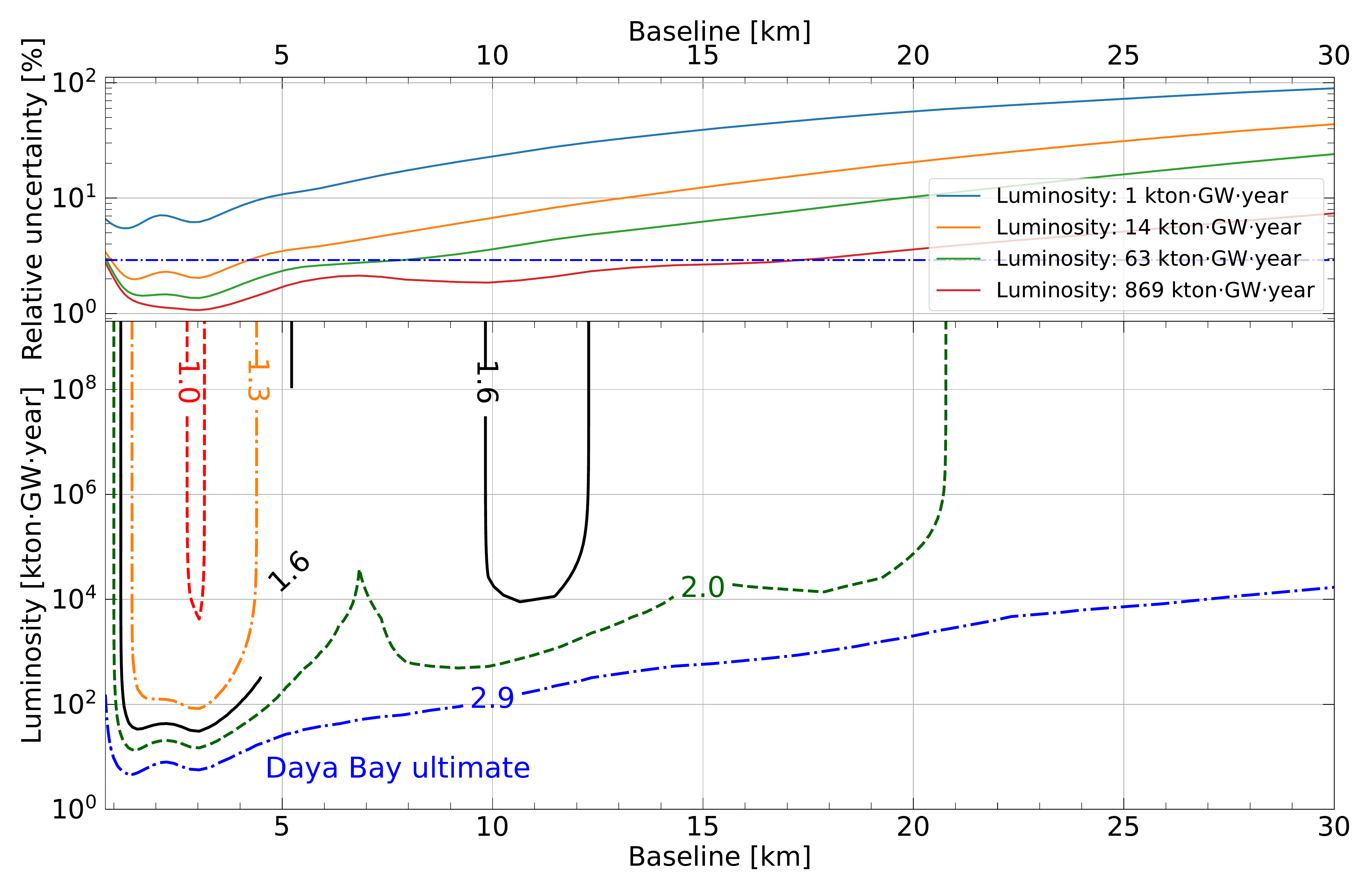}
  \caption{The 1$\sigma$ contour of the precision measurement sensitivity on \sinsq13  for different baselines and integrated luminosities. The top pad shows the sensitivity as a function of baselines for luminosities of 1, 14, 63, and 869~\ktonGWyear. The luminosities are arbitrarily selected to present the evolution of the baseline's preference.
    The optimal baseline shifts from $\sim 1.3$~km at low statistics to $\sim2.9$~km at high statistics. The ultimate precision of the Daya Bay experiment is also shown with a horizontal blue dash-dotted line.
    The bottom pad shows the required luminosity to reach the given precision, from 2.9\% to 1.0\% labeled on the curves, at different baselines.
    The spectral shape uncertainty is set to be 1\% for a bin width of 35~keV.
  }
  \label{fig:sensitivity_contour:baseline_luminosity:flat}
\end{figure}

With a 1\% spectral shape uncertainty, Fig.~\ref{fig:sensitivity_contour:baseline_luminosity:flat} shows the 1$\sigma$ contour of the precision measurement sensitivity on \sinsq13  for different baselines and integrated luminosities.
The optimal baseline for the most efficient \sinsq13 precision measurement varies with the increase of the integrated luminosity. The top pad of Fig.~\ref{fig:sensitivity_contour:baseline_luminosity:flat} shows the optimal baseline is about 1.3~km for low statistics $\lesssim 60$~\ktonGWyear~and gradually shifts to about 2.9~km as statistics increases. At $\sim 2.9$~km, the experiment could measure \sinsq13 to a precision of sub-percent level with a $\sim 10^4$~\ktonGWyear~luminosity.
Besides, we find that the precision is limited to $\sim 1$\% level by the spectral shape uncertainty, as the sensitivity will not be better with the increase of luminosity after $\sim 10^4$~\ktonGWyear. With such spectral shape uncertainty, the single detector at the first oscillation maximum of $\sim 1.8$~km would be impossible to measure \sinsq13 to the precision of a sub-percent level.

The baseline preference differs from those reactor experiments with identical near and far detectors~\cite{Huber:2003pm}, whose optimal baseline at about 1.8~km, the first oscillation maximum for reactor antineutrinos. The major difference is that the 3\% relative rate uncertainty is the dominant systematics for the single detector strategy, and it is suppressed to $\sim0.1$\% for the identical near and far detector configuration.
The oscillation parameter \sinsq13 characterizes the disappearance amplitude, which can be mimicked by the rate uncertainty nuisance parameter when the disappearance maximum is at the same energy as the measurable reactor antineutrino energy spectrum.
The offset of the oscillation maximum and unoscillated measurable antineutrino energy spectrum is shown in Fig.~\ref{fig:flux_oscillation}; the rate uncertainty is important for a single detector at low statistics.

As the statistics increase, the spectral shape distortion contributes more and more sensitivity since each energy bin has enough statistics to reflect the oscillation effect, and all bins share the same absolute rate uncertainty. The correlation among different bins thus suppresses the impact of the rate uncertainty at high statistics. Shifting the baseline from the rate oscillation maximum will help further reduce the impact of the rate uncertainty by offsetting the oscillation maximum energy bin from the peak of the reactor neutrino spectrum. At larger baselines, there are more oscillation cycles in the measured spectrum; thus, the error cancellation due to correlation among different bins is enhanced. Furthermore, multiple cycle measurement helps to reduce energy-correlated uncertainties. The optimal baseline comes from the balance of systematic uncertainty suppression and the statistics loss.

\begin{figure}[ht]
  \centering
  \includegraphics[width=0.8\textwidth]{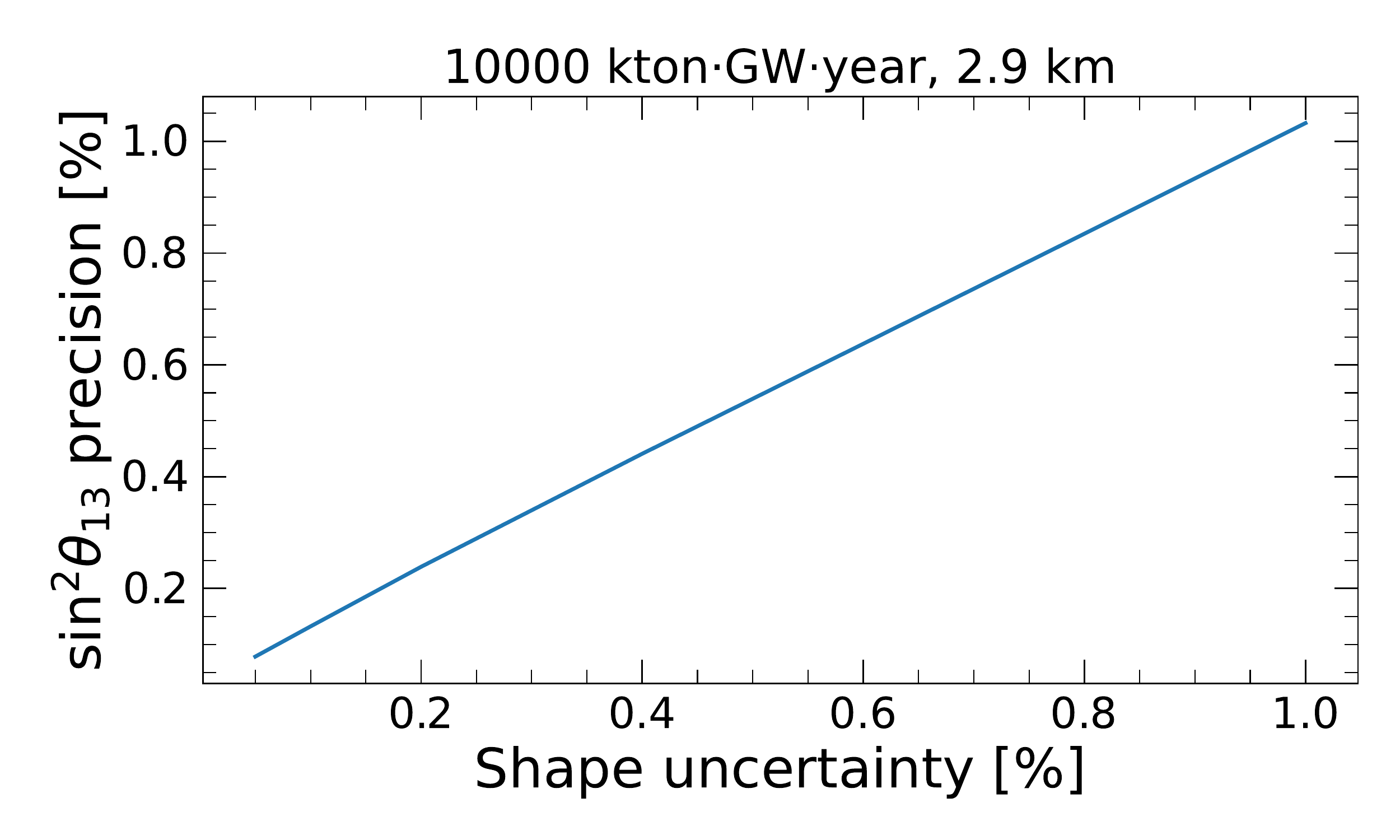}
  \caption{The \sinsq13 precision sensitivity for different input spectral shape uncertainty. The x-axis is the spectral relative shape uncertainty in \%. The baseline is 2.9~km, and the luminosity is $10^4$~\ktonGWyear. The \sinsq13 precision sensitivity is almost linearly proportional to the spectral shape uncertainty.}
  \label{fig:shape_err:amount:impact}
\end{figure}

With flat spectral shape uncertainty given by the future nuclear theory community's prediction, the optimal baseline keeps being $\sim 2.9$~km. We find that the optimal baseline is the same for different input shape uncertainty values (0.1\%, 0.5\%, 1\%, 2\%, 5\%), and with 1\% shape uncertainty, it would be the dominant systematics for the luminosity larger than 20~\ktonGWyear.
To verify the impact of the spectral shape uncertainty, here, we numerically calculate the sensitivity with different input shape uncertainties for an experiment with a luminosity of $10^4$~\ktonGWyear~at 2.9~km.
Fig.~\ref{fig:shape_err:amount:impact} shows the \sinsq13 precision measurement sensitivity for different input shape uncertainties. It is shown that the shape uncertainties are the bottleneck of the high precision measurement of \sinsq13.

\subsection{The impact of the shape uncertainty distribution}
\label{sec:sensitivity:shape_err}
\begin{figure}[htbp]
  \centering
  \includegraphics[width=\textwidth]{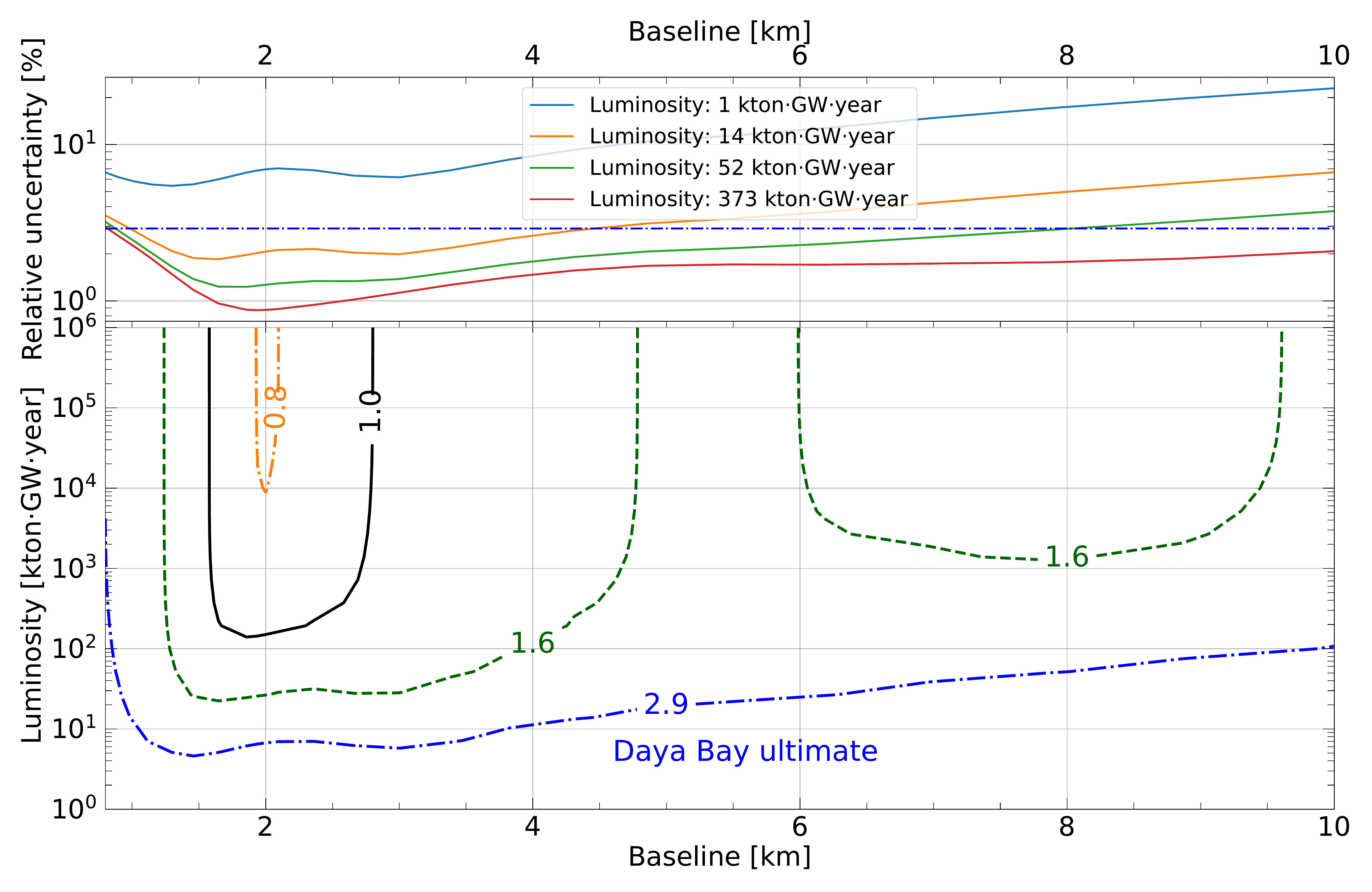}
  \caption{The 1$\sigma$ contour of the precision measurement sensitivity on \sinsq13  for different baselines and integrated luminosities. The top pad shows the sensitivity as a function of baselines for luminosities of 1, 14, 52, and 373~\ktonGWyear. The luminosities are arbitrarily selected to present the evolution of the baseline's preference.
    The optimal baseline shifts from $\sim 1.3$~km at low statistics to $\sim 2.0$~km at high statistics.
    The ultimate precision of the Daya Bay experiment is also shown with a horizontal blue dash-dotted line.
    The bottom pad shows the required luminosity to reach 0.8\% (orange dash-dotted line), 1.0\% (black line), 1.6\% (green dash line), and 2.9\% (dash-dotted blue line) precision at different baselines.
    The spectral shape uncertainty is constrained by the TAO experiment.
  }
  \label{fig:sensitivity_contour:baseline_luminosity:TAO}
\end{figure}

At large luminosity, the dominant uncertainty is the spectral shape uncertainty; thus, the optimal baseline highly depends on the energy distribution of the shape uncertainty.
In the above discussion, we use a 1\% spectral relative shape uncertainty.
In the near future, after the TAO experiment starts running, we can use the direct measurement of the reactor \nuebar~to constrain the spectral shape uncertainty.
The baseline preference will be different with such a spectral shape uncertainty setup.
Fig.~\ref{fig:sensitivity_contour:baseline_luminosity:TAO} shows the 1$\sigma$ contour of the precision measurement sensitivity on \sinsq13~using the TAO measurement as an external spectral shape uncertainty constraint.
It is shown that the optimal baseline is about 1.3~km for low statistics $\lesssim 10$~\ktonGWyear~and shifts to about 2.0~km as statistics increase. With the luminosity of $\sim 150$~\ktonGWyear, the experiment could measure \sinsq13 to a precision of sub-percent level.

With TAO-based spectral shape uncertainty, the required luminosity for measuring \sinsq13 to 1\% precision is less than that with the flat 1\% assumption. The reason is that the TAO-based uncertainty is $<1\%$ for $E_\nu\in (3,6)$~MeV, which is the peak of unoscillated reactor antineutrino and thus has the largest statistics.
At 2~km, the oscillation maximum is at the same energy as the peak of the unoscillated measurable reactor antineutrino energy spectrum. We can get more information for \sinsq13 oscillation with the spectral shape uncertainty constrained by the TAO experiment. In contrast, with the 1\% flat uncertainty model, the absolute spectral shape uncertainty is proportion to the statistics of the unoscillated measurable reactor antineutrino spectrum for all energy. Larger statistics bring larger absolute spectral shape uncertainty; thus, the optimal baseline in Fig.~\ref{fig:sensitivity_contour:baseline_luminosity:flat} is 2.9~km, offsetting the oscillation maximum from the spectrum's peak.

\subsection{Systematics breakdown and sub-percent precision}
\label{sec:sensitivity:systematics:breakdown}
\begin{figure}[htbp]
  \centering
  \includegraphics[width=\textwidth]{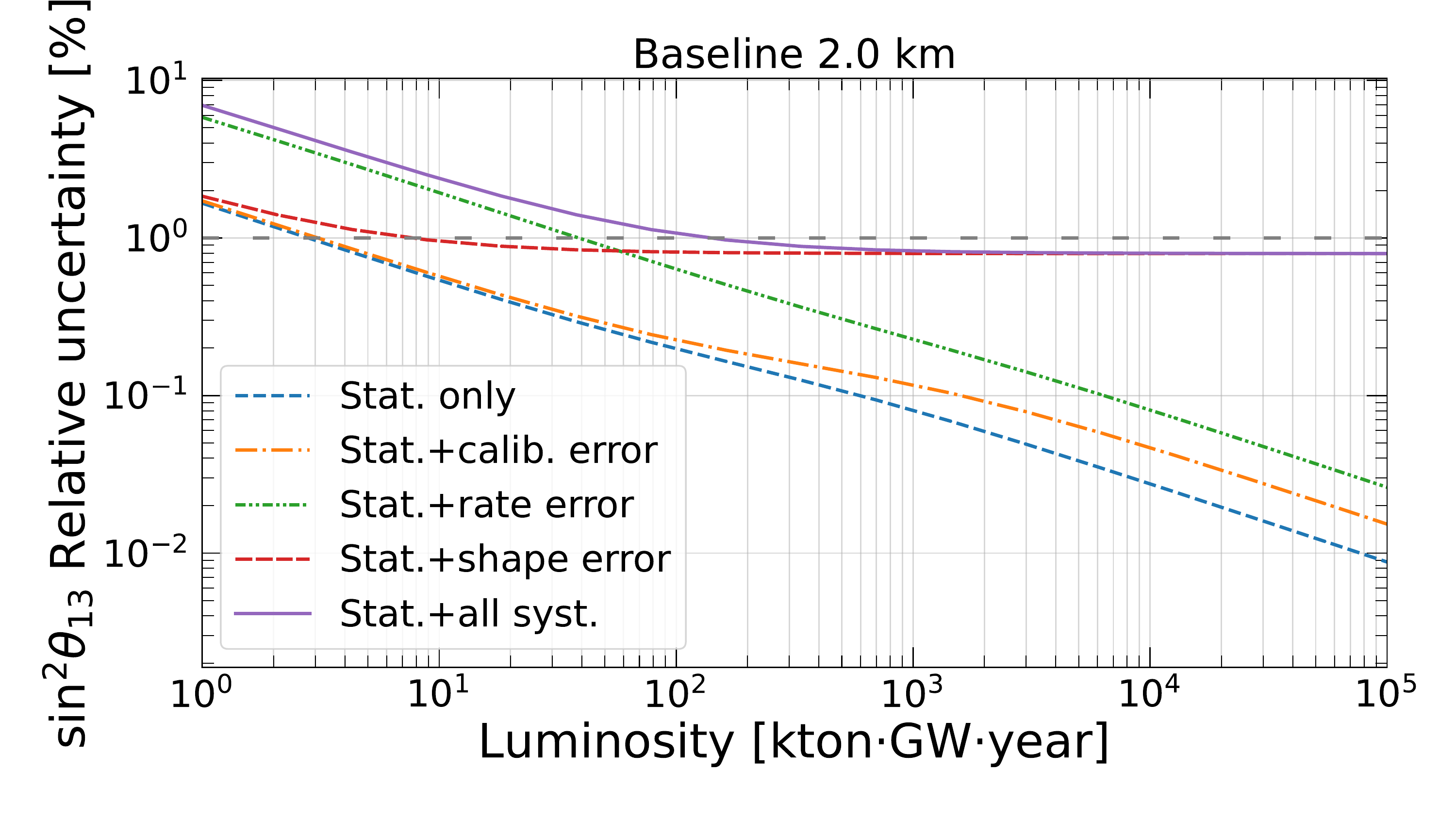}
  \caption{The \sinsq13 precision measurement sensitivity under different considerations of the systematic uncertainties as a function of integrated luminosity.
    The rate uncertainty is the dominant systematics at low statistics for the single detector strategy.
    With the integrated luminosity larger than $\sim 60$~\ktonGWyear, the shape uncertainty gradually dominates the precision measurement of \sinsq13. The ultimate precision sensitivity is approximately 0.8\%, limited by the spectral shape uncertainty.}
  \label{fig:sensitivity:syst.}
\end{figure}

Based on Fig.~\ref{fig:sensitivity_contour:baseline_luminosity:flat} and Fig.~\ref{fig:sensitivity_contour:baseline_luminosity:TAO}, the most efficient baseline for sub-percent measurement of \sinsq13 is about 2.0~km. Thus, we set the nominal baseline to be 2.0~km and study the impact of different systematic uncertainties. Fig.~\ref{fig:sensitivity:syst.} shows the breakdown of the statistical and systematic uncertainties for the precision measurement sensitivity of \sinsq13 at different luminosities.
It helps us to identify the most important systematic uncertainties at different luminosities. The rate uncertainty would be the dominant systematics at low luminosity $\lesssim$ 60~\ktonGWyear; then, the shape uncertainty would be dominant. The impact of energy scale calibration uncertainty is negligible at low statistics and would have a minor impact as luminosity increases. The shape uncertainty would totally dominate the \sinsq13 precision measurement with luminosity larger than $\sim 10^3$~\ktonGWyear.

With the integrated luminosity of about 150~\ktonGWyear, the experiment can measure \sinsq13 to the sub-percent level precision. One feasible design is to install a 10\% energy resolution, 4~kton liquid scintillator detector near a reactor complex like the Taishan reactor, whose thermal power is about 9.2 GW. With such a setup,
consider a typical $>80$\% signal selection efficiency and 11/12 reactor duty cycle~\cite{DayaBay:2016ggj},
the experiment could measure \sinsq13 to the sub-percent precision level within four years.

The spectral shape uncertainty is crucial for the \sinsq13 sub-percent precision measurement. The shape uncertainties given by different methods have either the close energy distribution as the TAO-based model (direct measurement) or the flat 1\% model (theoretical calculation). We calculate the \sinsq13 precision sensitivity by multiplying the TAO uncertainty curve with different factors (>1). The optimal baseline is the same for different factors. The precision sensitivity has a similar dependence on shape uncertainty as Fig.~\ref{fig:shape_err:amount:impact}. The results show that with the spectral shape uncertainty $1.3$ times as large as the TAO uncertainty curve, the single detector strategy's ultimate precision sensitivity would be larger than 1\%. The 30\% buffer of the shape uncertainty provides ample space for our future experiment on measuring \sinsq13 to a sub-percent precision.

\subsection{The impact of the oscillation parameters}
\label{sec:sensitivity:oscillation}
\begin{figure}[htbp]
  \centering
  \includegraphics[width=\textwidth]{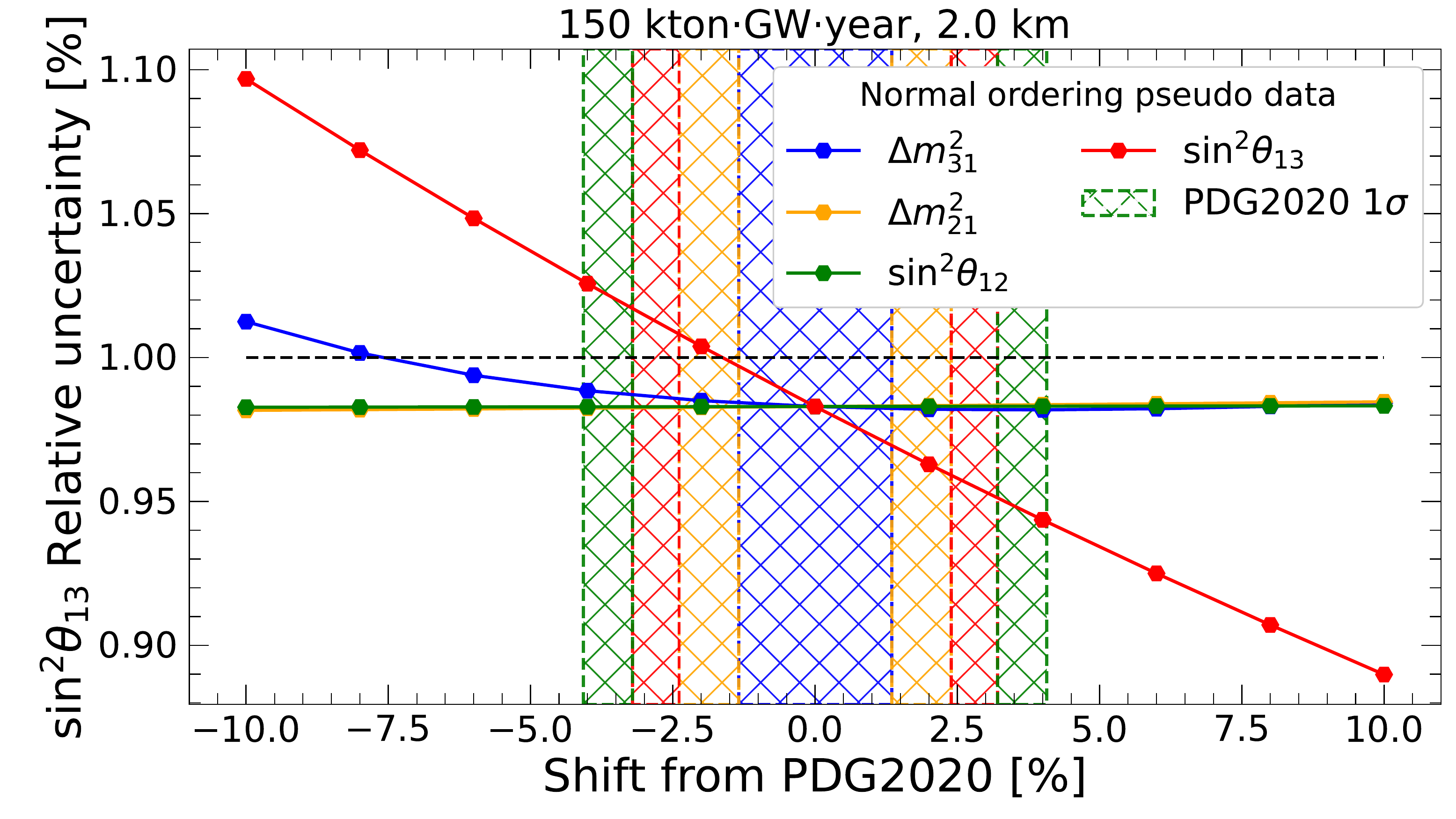}
  \caption{The \sinsq13 precision measurement sensitivity for the Asimov pseudo data generated with the oscillation parameters shifted from PDG2020~\cite{ParticleDataGroup:2020ssz}. The green and orange lines are on the top of each other since the roles of \sinsq12 and \dmsq21 true values are almost the same for the \sinsq13 measurement. The shadow area represents the 1$\sigma$ region given in PDG2020. The normal and inverted ordering hypotheses yield almost the same results. It can be seen that the true value of \sinsq13 has more impact than all other oscillation parameters. }
  \label{fig:sensitivity:osci:NMO}
\end{figure}

The optimal baseline ($\sim 2.0$~km) we propose in this work depends on the oscillation parameters' central values and the reactor antineutrino energy spectrum. Thus, we generate several Asimov data sets with different oscillation parameter values and mass orderings to assess the impact. The results using the nominal luminosity 150~\ktonGWyear~is shown in Fig.~\ref{fig:sensitivity:osci:NMO}, where we always update \dmsq32 correspondingly assuming \dmsq32=\dmsq31$-$\dmsq21.

It can be seen that the \sinsq13 measurement is almost independent of the true values of the solar oscillation parameters \sinsq12 and \dmsq21 at such baseline. The sensitivity weakly depends on the true value of \dmsq31.
As we expected, the relative precision sensitivity is almost inversely linear proportional to the true value of \sinsq13. This dependence shows that the ability of the detector to locate the absolute value of \sinsq13 is robust.

The true value of \sinsq13 is important for the sub-percent relative precision measurement of \sinsq13; if the true value is smaller, the detector will have to take more data for the sub-percent precision measurement. The central value estimated by PDG2020~\cite{ParticleDataGroup:2020ssz} is 0.0218, and recently the results from global fit groups yield that the central value is 0.0223 ($\sim 2$\% larger)~\cite{Gonzalez-Garcia:2021dve,Capozzi:2021fjo, deSalas:2020pgw}. With a larger true value, the experiment can measure \sinsq13 to the sub-percent precision level with fewer statistics.

In Sec.~\ref{sec:detection-statistics:statistics}, we constrain \sinsq12, \dmsq21, and \dmsq31 using the PDG2020 central values and 1$\sigma$ uncertainties by adding a pull term defined in Eq.~\eqref{eq:pull:oscillation:parameters},
\begin{equation}
  \chi^2_{\rm pull}\equiv \sum_\theta \frac{\left( \theta-\theta_0\right)^2}{\sigma^2_\theta},
  \label{eq:pull:oscillation:parameters}
\end{equation}
where $\theta$ refers to the oscillation parameters, including \sinsq12, \dmsq21, and \dmsq31. $\theta_0$ and $\sigma_\theta$ are the central values and 1$\sigma$ uncertainties. With the nominal setup, we find that the \sinsq13 precision sensitivity keeps almost unchanged (relative difference within 0.2\%) whether we constrain, fix, or free \dmsq31. Actually, the experiment can also measure \dmsq31 to the precision of $\sim 0.6$\% with 150~\ktonGWyear\ luminosity.
As a future experiment, we can anticipate the future external information of the oscillation parameters. With the \sinsq12, \dmsq21, and \dmsq31 constraints from the projected relative precision listed in Table~\ref{tab:current:knowledge}, the \sinsq13 precision sensitivities would be slightly better (relative difference $\sim$0.5\%).
When we fix all other oscillation parameters, \sinsq12, \dmsq21, and \dmsq31,  the \sinsq13 precision sensitivity would be almost the same as the projected relative precision (relative difference within 0.1\%). The experiment can measure \sinsq13 to the sub-percent precision level with other oscillation parameters fixed or constrained with external information. However, since this detector is not designed for \sinsq12 and \dmsq21 measurements, when we free all other oscillation parameters, the nominal sensitivity reduces from 1\% to $\sim 2.6$\%. Thus, for a \sinsq13 high precision measurement experiment, using external information could help accomplish the major physics goal.

\subsection{The impact of the reactor antineutrino anomaly and excess}
\label{sec:sensitivity:RAA}
As shown in Fig.~\ref{fig:flux_oscillation}, the offset of the optimal baseline depends on the unoscillated measurable reactor \nuebar~energy spectrum. In Sec.~\ref{sec:detection-statistics:detection}, we use the Huber-Mueller model~\cite{Huber:2011wv, Mueller:2011nm} as the nominal flux model to predict the observed reactor antineutrino energy spectrum. 
However, a $\sim 6$\% absolute flux deficit compared to the Huber-Mueller model is observed by many reactor neutrino experiments~\cite{Declais:1994ma, DayaBay:2018heb, DANSS:2018fnn, DoubleChooz:2019qbj, NEOS:2016wee, NEUTRINO-4:2018huq, PROSPECT:2020sxr, RENO:2020dxd, STEREO:2020hup}, with the leading evidence provided by the Daya Bay~\cite{DayaBay:2018heb}, Double Chooz~\cite{DoubleChooz:2019qbj}, and RENO~\cite{RENO:2020dxd} experiments.
This deficit is the so-called reactor antineutrino anomaly. Besides, the experiments also find an excess around 5~MeV on the spectral shape called ``5~MeV excess''~\cite{Athar:2021xsd}.

We can use the flux measured by previous reactor neutrino experiments to explore the impact of reactor antineutrino anomaly and excess. This work employs the unfolded isotope flux from the Daya Bay experiment given in Ref.~\cite{DayaBay:2021dqj}, which naturally includes both effects. In the interest of studying the flux model dependence of the sensitivity, we switch the flux calculation of all isotopes in Eq.~\eqref{eq:spec-prediction} from $\sum_i f_{i}\phi_i(E_{\bar{\nu}})$ to Eq.~(8) of Ref.~\cite{DayaBay:2021dqj}. With all the other setups, including systematics, same as in Sec.~\ref{sec:detection-statistics}, Table~\ref{tab:flux:model:syst} gives the 1$\sigma$ precision sensitivity of \sinsq13 with different isotope flux models.
\begin{table}[htbp]
  \centering
  \begin{tabular}{lcc}
    \toprule
    \sinsq13 1$\sigma$ uncertainty (\%) & Huber+Mueller~\cite{Huber:2011wv,Mueller:2011nm} & Daya Bay unfolded~\cite{DayaBay:2021dqj} \\
    \midrule
    Stat.                               & 0.170                                            & 0.173                                    \\
    Stat.+rate error                    & 0.524                                            & 0.535                                    \\
    Stat.+calib. error                  & 0.198                                            & 0.200                                    \\
    Stat.+shape error                   & 0.807                                            & 0.808                                    \\
    Stat.+all syst.                     & 0.983                                            & 0.986                                    \\
    \bottomrule
  \end{tabular}%
  \caption{The \sinsq13 1$\sigma$ precision sensitivity under different consideration of the uncertainties with the Huber-Mueller flux model and the Daya Bay unfolded flux model. The latter naturally includes the reactor antineutrino anomaly and the ``5~MeV excess''. The 6\% deficit reduces the statistics and the ``5~MeV excess'' slightly benefits the \sinsq13 precision measurement. All results are for the detector installed at 2.0~km from the reactor and 150~\ktonGWyear\ luminosity.}
  \label{tab:flux:model:syst}%
\end{table}%

It can be seen that the reactor antineutrino anomaly and ``5~MeV'' excess have a minor impact on the sub-percent precision measurement of \sinsq13. The latter effect means there are more events at the neutrino energy around 6~MeV, which helps the measurement by providing more statistics at this energy. As shown in Fig.~\ref{fig:flux_oscillation}, more statistics around 6~MeV slightly improve the sensitivity as shown in Table~\ref{tab:flux:model:syst}.
The Double Chooz experiment also demonstrates a similar conclusion using its single-detector configuration data~\cite{DoubleChooz:2019qbj}.
With the existing measurements and the coming high precision measurement from TAO as input, the reactor antineutrino anomalies will not bias the sub-percent measurement of \sinsq13.

\subsection{The impact of the detector performance}
\label{sec:sensitivity:detector}
In Sec.~\ref{sec:detection-statistics:detection}, we assume the detector's energy resolution to be 10\%.
The high energy resolution is usually expensive and technically challenging for a large-volume liquid scintillator detector.
\begin{figure}[htbp]
  \centering
  \includegraphics[width=\textwidth]{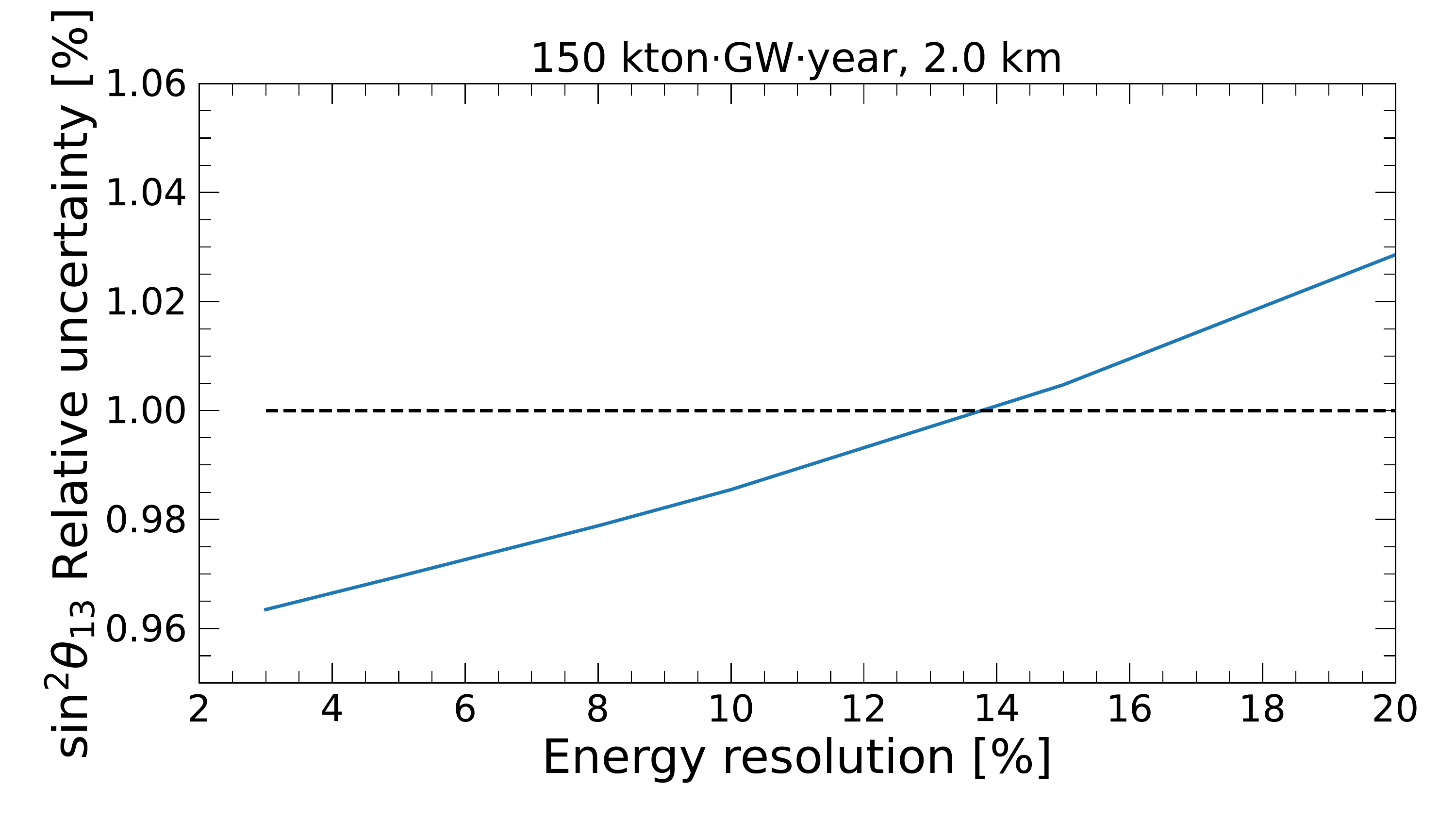}
  \caption{The \sinsq13 1$\sigma$ precision sensitivity for the experiment with different prompt energy resolutions. The \sinsq13 precision sensitivity's relative variation would be within 5\% for the energy resolution from $3\%/\sqrt{E({\rm MeV})}$ to $15\%/\sqrt{E({\rm MeV})}$.}
  \label{fig:sensitivity:resolution}
\end{figure}
Thus, with the nominal setup, we study the \sinsq13 precision measurement sensitivity for a detector with different energy resolutions. As shown in Fig.~\ref{fig:sensitivity:resolution}, the relative variation of sensitivity would be within 5\% for the energy resolution from $3\%/\sqrt{E({\rm MeV})}$ to $15\%/\sqrt{E({\rm MeV})}$. In general, the energy resolution is not a key factor for \sinsq13 precision measurement at the baseline of $\sim 2.0$~km. A 10\% energy resolution is sufficient for measuring \sinsq13 to the sub-percent precision level.

As a detector with volume at the kton level, we assume that the energy scale calibration's high accuracy  (uncertainty $<$0.5\%) can be achieved as the Daya Bay experiment~\cite{DayaBay:2019fje} and the prospects of the JUNO experiment~\cite{JUNO:2020xtj}. Nonetheless, even if the energy scale uncertainty is at a 1\% level, the \sinsq13 precision measurement sensitivity keeps almost the same (relative difference $<$0.1\%). The result is consistent with the calibration uncertainty contribution we observe in Fig.~\ref{fig:sensitivity:syst.}.

\section{Conclusion}
\label{sec:conclusion}
For measuring \sinsq13 to the sub-percent level precision, the crucial requirements are the statistics, the baseline, and the control of the spectral shape uncertainty.
We perform a numerical calculation of the \sinsq13 precision measurement sensitivity and find that the optimal baselines for a single liquid scintillator detector setup are different from the identical near and far detectors setup. The latter setup can suppress the rate uncertainties by near-far relative measurement, and the optimal baseline is about 1.8~km. The optimal baseline for the former setup is about 1.3~km at low luminosities $\lesssim 10$~\ktonGWyear~as the dominant systematics is the rate uncertainty. For larger statistics, as the shape uncertainty becomes dominant, the optimal baseline shifts to about 2.0~km and keeps being so for the integrated luminosity up to $10^6$~\ktonGWyear.
The reason is that \sinsq13 characterizes the disappearance amplitude; thus, the rate uncertainty plays an important role in the measurement when the disappearance maximum is at the peak of the unoscillated antineutrino energy spectrum. For a single detector experiment with large rate uncertainty, the optimal baselines shift from the baseline of the maximum rate oscillation.

With the spectral shape uncertainty constrained by the TAO experiment, a single liquid scintillator detector at the baseline of $\sim 2.0$~km with a JUNO-like overburden could measure \sinsq13 to the sub-percent precision level within 150~\ktonGWyear~integrated luminosity.
The energy resolution is not a key factor for an experiment at several kilometers' baselines. Thus, we propose to install a single 4~kton, 10\% energy resolution detector at $\sim 2.0$~km from a 9.2~GW reactor complex like the Taishan reactor. The experiment with such a setup could measure \sinsq13 to the sub-percent precision level within four years.
Various factors that may increase or decrease the sensitivity are discussed; the dominant factors are the reactor antineutrino spectral shape uncertainty and the \sinsq13 true value.
With the flat relative spectral shape uncertainty given by the future nuclear theory community's prediction, the optimal baseline is 2.9~km at larger luminosities.
The relative shape uncertainty is the same for different energies in this model; thus, the optimal baseline is further shifted to offset the oscillation maximum of 1.8~km at the peak of the reactor neutrino spectrum.
The detector performances on the energy resolution and energy scale uncertainty have minor impacts on the sensitivity. Since we set a 0.1\% spectral shape uncertainty of the background subtraction, the experiment should have good control of the background, such as the natural radioactivity and cosmogenic backgrounds.

\section*{Acknowledgements}
This work was supported by the National Key R\&D Program of China (2018YFA0404101).

\bibliographystyle{JHEP}
\bibliography{SubPercentTheta13WithReactor}
\end{document}